\documentstyle[onecolumn,epsf,rotate]{mn}
\begin{document}

\onecolumn

\title[]
      {Non-adiabatic tidal forcing of a massive, uniformly rotating
star III: Asymptotic treatment for low frequencies in the inertial regime}

\author[J.C.~Papaloizou and G.J.~Savonije]
       { J.C.~Papaloizou $^{1}$ and G.J.~Savonije $^{2}$\\
	$^{1}$ Astronomy Unit, School of Mathematical Sciences, Queen Mary 
        and Westfield College, University of London, Mile End Road, \\
        London E1 4NS, UK \\
	$^{2}$ Astronomical Institute `Anton Pannekoek', University of Amsterdam
and Centre for High Energy Astrophysics (CHEAF),\\ Kruislaan 403,
1098~SJ~Amsterdam, The Netherlands}

\maketitle


\newcommand{\bra}[1]{\left( #1 \right)}
\newcommand{\brac}[1]{\left\{ #1 \right\}}
\newcommand{\brasq}[1]{\left[ #1 \right]}

\newcommand{\pdrv}[2]{{{\partial #1}\over {\partial #2}}}
\newcommand{\pddrv}[2]{{{\partial^2 #1}\over {{\partial #2}^2}}}
\newcommand{\drv}[2] {{{d #1}\over {d #2}}}
\newcommand{\deriv} [1] {{d \over {d #1}}}
\newcommand{\vc}[1] {\mbox{\boldmath$#1$}}
\newcommand{\el}{{\it l}}
\newcommand{\ds}{\displaystyle\strut}
\newcommand{\andd}{\ {\rm and} \ }
\newcommand{\etal}{{\em et al.} }
\newcommand{\half}{{1 \over 2}}
\newcommand{\eq}{\begin{equation}}
\newcommand{\ee}{\end{equation}}
\newcommand{\new}{\item [\ ]}
\newcommand{\fr}{{\cal I}}

\begin{abstract}

\noindent We describe a generalization of the asymptotic calculation
of the tidal torques experienced by a massive star as a result of a companion
in circular orbit originally considered by Zahn (1975,1977) 
 to the case of a rotating star when the forcing frequency is
small and in the inertial regime, that is it is less than twice the rotation
frequency in magnitude. The results confirm the presence of a strong toroidal
mode resonance feature for retrograde forcing and also, with a simple
description of the convective core, the presence of some core inertial mode
features in the response. These were found numerically by Savonije and
Papaloizou (astro-ph/9706186).  \end{abstract} 


\section{Introduction}
In two recent papers, ( Savonije, Papaloizou and Alberts, 1995: SPA,
Savonije and Papaloizou, 1997: SP) we studied the response of a uniformly
rotating massive star to the tidal forcing due to a companion in a circular
orbit (e.g. a massive X-ray binary). To make the problem tractable, centrifugal
forces were neglected, allowing a spherically symmetric equilibrium, 
but Coriolis forces were retained allowing normal modes governed by rotation
to enter into the response. This problem is complex because  the dense
spectrum of normal modes causes excitation of short wavelength responses
which lead to numerical difficulties. As a result the analysis of SPA
could not be extended into the inertial regime in which the forcing frequency
is less than  twice the rotation frequency in magnitude. In SP numerical results
appropriate to the inertial regime for modest stellar rotation rates have been
obtained through the introduction of a viscosity which provides  numerical
damping of the shortest wavelengths in the response in convective regions.

\noindent The first calculations of the tidal torques
experienced by a non-rotating massive star as a result of a companion
in a circular orbit were performed by Zahn (1975,1977) who developed an
asymptotic approach valid for low forcing frequencies. 
Papaloizou and Savonije (1984, 1985) found that this could give reasonable results
provided the star was unevolved with the convective core not being too small
in size.

\noindent  In this paper we investigate a generalization of the asymptotic
calculation of the tidal torques  to the case of a rotating star when the
forcing frequency is small in magnitude and in the inertial regime. Such an
analysis is of interest in order to complement the numerical work which runs
into difficulties because of the short wavelength nature of the response at low
frequencies. We find that some of the phenomena found by SP, in particular the
strong toroidal mode resonance, is present in the asymptotic analysis  as well
and persists in
the low frequency limit. This may be important for retrograde forcing leading
to a more rapid synchronization of spin and orbit than would be expected from
consideration of non-rotating stars.

\noindent In section 2 we give the basic equations for a uniformly rotating
star subject to tidal perturbation due to a companion in circular orbit,
formulating a single equation giving the response of the adiabatic interior.
In section 3 we consider the response of the convective core, indicating how
inertial modes may be excited there. We also consider the radiative exterior
indicating how use of the traditional approximation leads to a separable
problem for excited generalized $g$ and $r$ modes in that region. 
In section 4 we consider the WKB approximation for the radiative region with
outgoing wave conditions
and how toroidal mode resonances lead to a global response. We go on to
evaluate the action of the tidal torque in producing an angular momentum
flux through wave excitation in section 5. In section 6 we describe numerical results while
in section 7 we summarise the conclusions.

 \section{Basic Equatiions}
We consider a uniformly rotating massive secondary star with mass $M_s$ and
radius $R_s$. We assume the angular velocity of rotation $\vec{\Omega_s}$ to
be much smaller than the break up speed, i.e. $(\Omega_s/\Omega_c)^2\ll 1$,
with $\Omega_c^2=GM_s/R_s^3$, so that effects of centrifugal distortion
($\propto \Omega_s^2$) may be neglected in first approximation.

\noindent We use spherical coordinates
($r,\theta,\varphi)$, with origin at the secondary's centre, whereby
$\theta=0 $ corresponds to its rotation axis which we assume to be
parallel to the orbital angular momentum vector. However,  we take the
coordinates to be  rotating with the secondary.

\noindent  The linearized
hydrodynamic equations governing the non-adiabatic response of the uniformly
rotating star to the perturbing potential $\Phi_T$ are the equation of motion

\eq \pdrv{ {\bf v'}}{t} + 2\Omega_s{\bf k}\times {\bf v'}
=-\frac{1}{\rho} \nabla P' + \frac{\rho'}{\rho^2} \nabla P - \nabla \Phi_T,
\label{eqmot} \ee
and the equation of continuity

\eq  \pdrv{\rho' }{t} +  \nabla\cdot(\rho {\bf v'})=0
 \label{eqe} \ee

\noindent Here the velocity perturbation is ${\bf v'}$ and the  pressure  and
density perturbations are $P'$ and  $\rho' $ respectively.
The unit vector along the rotation axis is denoted by ${\bf k}.$

\noindent We consider a close binary system in which the orbit is circular with angular
velocity $\omega$ and orbital separation $D$.  The dominant tidal term of
the primary's perturbing potential is then given by the real part of:

\eq \Phi_T= f r^2 \, P^2_2 (\mu) \,\,\exp(2 i [\omega_f t - \varphi ])\equiv
\Phi_{T0} \, \exp(2 i [\omega_f t - \varphi ])
\label{eqpot} \ee
where $M_p$ is the companion's mass, $\omega_f = \omega -\Omega_s$ is the
relative orbital frequency, as seen in the rotating frame, $\mu=cos\theta$,
$P^2_2(\mu)$ is the associated Legendre polynomial for $l=|m|=2$ and $$ f=
-\frac{G M_p}{4 D^3}.$$

\noindent  For simplicity we have adopted the
Cowling approximation, i.e. we have neglected perturbations to the 
secondary's gravitational
potential caused by  tidal distortion. This approximation is
reasonable because  of the high central condensation.

\noindent In this paper we adopt a simple model for calculating the tidal
response in which the angular momentum exchange between the rotating star
and the binary orbit occurs through angular momentum carying waves
that are excited in the region of the convective core boundary and the
radiative layers in its neighbourhood. It is supposed that these waves are 
subsequently dissipated with the consequent angular momentum exchange
(see Papaloizou and Savonije  1984,1985, and Goldreich and Nicholson 1989).
In order to analyse this model, only the interior regions of the star
need to be considered. Then the perturbations are (almost) adiabatic such that

\eq  \pdrv{P' }{t} - \frac{\Gamma P}{\rho} \pdrv{\rho' }{t}
+{\bf v'}\cdot\left(\nabla P - \frac{\Gamma P}{\rho} \nabla \rho \right) =0.
 \label{eade} \ee
For simplicity in what follows below we shall take $\Gamma$
to be constant and equal to $5/3.$ Then (\ref{eqe}) and (\ref{eade}) can be
combined to give \eq  \pdrv{P' }{t}= -\Gamma P^{(1-1/\Gamma)}\nabla\cdot(
P^{1/\Gamma} {\bf v'}) \label{ePe}.\ee

\subsection{Reduction of the Response Equations}
The linear  response to the tidal potential (\ref{eqpot}) is such that
all perturbations have a $\varphi$ and $t$ dependence through a factor
$\exp(2 i [\omega_f t - \varphi ]).$ But note  that, due to the occurrence of
the Coriolis terms in the equations, the solution is no longer separable in
$\theta$, as it is for non-rotating stars. Equations
 (\ref{eqmot}) and (\ref{eqe}) give
for the components of the displacement $\vc{\xi}=(i\sigma)^{-1}{\bf v}'$:

\eq -\sigma^2\xi_r-2i\Omega_s\sigma\sin\theta\xi_{\varphi}=
-K{{\partial W}\over \partial r}-(\xi_r+\Phi_{T}/g){\cal A},\label{M1}\ee
\eq -\sigma^2\xi_{\theta}-2i\Omega_s\sigma\cos\theta\xi_{\varphi}=
-{K\over r}{{\partial W}\over \partial \theta},\label{M2}\ee
\eq -\sigma^2\xi_{\varphi}+
2i\Omega_s\sigma\left(\cos\theta\xi_{\theta}+\sin\theta\xi_{r}\right)=
-{K\over r\sin\theta}{{\partial W}\over \partial \varphi}.\label{M3}\ee
Here $W= (P'+\rho \Phi_{T})/P^{1/\Gamma},$ and $K=P^{1/\Gamma}/\rho.$
The square of the Brunt-Vaisala frequency is given by
$${\cal A}=
\frac{\nabla P}{\rho}\cdot\left(\frac{\nabla
\rho}{\rho} -\frac{\nabla P}{\Gamma P} \right),$$ the local acceleration
due to gravity is given by
$g=-\rho^{-1}(dP/dr),$ and the forcing frequency $\sigma =2\omega_f.$
We can use equations (\ref{M1}) - (\ref{M3}) to express $\vc{\xi}$ in terms of
$W.$ Using the notation $(\xi_r,\xi_{\theta},\xi_{\varphi}) \equiv 
(\xi_1,\xi_2,\xi_3),$ $ (h_1,h_2,h_3) \equiv  (1, 1/r ,1/(r\sin\theta)), $and
$(x_1,x_2,x_3) \equiv (r,\theta ,\varphi),$ we may write (using the summation
convention) \eq \xi_i = \frac{A_{ij}}{\Delta}\left ( K h_j{\partial W
\over \partial x_j} +\delta_{j1}\frac{\Phi_{T}}{g}{\cal A}\right).
\label{xii}\ee The components of the Hermitian ( for real $\sigma$ ) matrix
$\left [ A \right ]$ are given by......
$$A_{11}=-\sigma^4+4\sigma^2\Omega^2_s\cos^2\theta,\,\,\,\, A_{12} =
-4\Omega_s^2\sigma^2 \sin\theta\cos\theta,\,\,\,\, A_{13}=2i\Omega_s\sigma^3\sin\theta
$$ $$A_{22}=\sigma^2(-\sigma^2 + {\cal A})+4\sigma^2\Omega^2_s\sin^2\theta,\,\,\,\,
A_{23}=-2i\Omega_s \sigma\cos\theta(-\sigma^2 + {\cal A}), \,\,\,\,
A_{33}=\sigma^2(-\sigma^2 + {\cal A}),$$ with the unspecified components being
given by the Hermitian condition assuming $\sigma$ to be real and 
$$\Delta =(-\sigma^2 + {\cal
A})(\sigma^4-4\sigma^2\Omega^2_s\cos^2\theta)
+4\sigma^4\Omega_s^2\sin^2\theta.$$
Using the above expression for the components of the displacement
in (\ref{ePe}) gives a single second order partial differential
equation for $W$ in the form

\eq P^{1/\Gamma}W - \rho \Phi_{T}= -\Gamma
P^{(1-1/\Gamma)}\frac{h_i}{q_i}\frac{\partial}{\partial x_i}\left( P^{1/\Gamma} 
q_i\frac{A_{ij}}{\Delta}\left ( K h_j{\partial W
\over \partial x_j} +\delta_{j1}\frac{\Phi_{T}}{g}{\cal A}\right)\right)
\label{eWe},\ee
where $(q_1,q_2,q_3)=(r^2,\sin\theta, 1).$

\noindent As the $\varphi$ and $t$ dependence of $W$ is through a separable factor
$\exp(im(\omega_f t-\varphi)),$ with $m \equiv 2,$  after the replacement  
${\partial\over \partial \varphi}
\rightarrow -im$ and $\Phi_{T}\rightarrow \Phi_{T0},$
 equation (\ref{eWe})  becomes  a second order
partial differential equation of mixed type for $W$ as a function of $r$ and
$\theta.$ It is hyperbolic whenever a real wave vector $(k_1,k_2,0)$ exists
such that $A_{ij}k_ik_j=0.$ This condition is equivalent to the requirement that
$\sigma$ satisfies the local dispersion relation (see Tassoul 1978, SPA)
\eq\sigma^2 = {k_2^2{\cal A}+4\Omega^2_s(k_1\cos\theta - k_2\sin\theta)^2
\over(k_1^2+k_2^2)}\label{dispr}.\ee  When this can be satisfied and the
boundary conditions  are non-dissipative,
one expects a dense spectrum of normal modes and a
singular response to forcing (SPA). This problem is avoided here by, from now on, 
allowing $\sigma$ to have a small negative imaginary part. This latter Landau
prescription corresponds to the forcing potential being slowly switched on
at time $t=-\infty.$  This also leads  naturally to the selection of a
predominantly outgoing
wave
 boundary condition,  corresponding to the
physical situation where the surface regions of the star are assumed  to be
highly dissipative with little or no wave reflection  occurring from them.

\section{Asymptotic treatment for low frequencies}
Because of lack of separability, the solution of equation (\ref{eWe})
cannot be undertaken analytically in general. However, progress can be made
if we assume that the square of the forcing frequency $\sigma^2$ is small
compared to the $G\rho,$ the square of the inverse of the local dynamical 
timescale. We shall be specially interested in  the inertial regime and so we
shall assume
$\sigma^2/\Omega_s^2$ is comparable to unity.
\subsection{The radiative zone}
In a radiative region ${\cal A}$ is non zero and positive and for low
frequencies $\sigma^2 /{\cal A}$ can be considered to be small,
apart possibly from a region in the vicinity of the interface between
the convective core and surrounding radiative zone.
However, what follows below can be shown to be valid also in such a region.
We write equation (\ref{eWe}) in the form
\eq P^{1/\Gamma}W +\Gamma
P^{(1-1/\Gamma)}\frac{h_i}{q_i}\frac{\partial}{\partial x_i}\left( P^{1/\Gamma} 
q_i\frac{A_{ij}}{\Delta} K h_j{\partial W
\over \partial x_j}\right)=S
\label{eWeS},\ee
where we have grouped the forcing terms involving $ \Phi_{T0}$ together
on the right hand side of (\ref{eWeS}) in $S.$ Considering that 
$\epsilon \equiv \sigma^2 /{\cal A}$ defines a small parameter, we may 
attempt to find solutions by expanding
the left hand side of (\ref{eWeS}) to first order in this parameter.
We also allow the solution to vary rapidly with radius such that
$\frac{\partial}{\partial r} = O(\epsilon^{-1/2}){1\over r}.$ It is further 
assumed that the angular variation is much less than the radial variation
which is the situation for the expected excited low frequency $g$ modes.
Then expansion of (\ref{eWeS}) in powers of $\epsilon$ gives to zero order

\eq P^{1/\Gamma}W\sigma^2 -{\sigma^2 \Gamma P^{(1-1/\Gamma)}\over r^2}
\frac{\partial}{\partial r}\left(\frac{r^2 K P^{1/\Gamma}}{{\cal A}} 
 {\partial W\over \partial r}\right) +\frac{K \Gamma 
P}{r^2}O_{\perp} (W) =\sigma^2 S
\label{eWeb},\ee
where the operator $O_{\perp}$ is defined through
\eq O_{\perp}(W)= -\frac{\partial Q}{\partial \mu}
+\frac{2m\Omega_s \mu  Q}{\sigma (1-\mu^2)}-\frac{m^2 W}{(1-\mu^2)},\ee
with $Q$ being related to $W$ through
\eq Q\left(1- \frac{4\Omega_s^2 \mu^2 }{\sigma^2 }\right)=
-(1-\mu^2)\frac{\partial W}{\partial \mu}
-\frac{2m\Omega_s\mu  W}{\sigma }.\ee
\subsection{The convective core}
We idealize the convective core to be a region in which
 ${\cal A}=0$ interior to the boundary, where we suppose  ${\cal A}=0,$
but  $\nabla {\cal A}\ne 0.$
When ${\cal A}=0,$  we cannot make an expansion
based on the smallness of $\sigma^2 /{\cal A}.$ In this case we must 
retain (\ref{eWeS}). However, $S$ takes the simple form $S=\rho \Phi_{T0},$
and thus (\ref{eWeS}) becomes
\eq P^{1/\Gamma}W +\Gamma
P^{(1-1/\Gamma)}\frac{h_i}{q_i}\frac{\partial}{\partial x_i}\left( P^{1/\Gamma} 
q_i\frac{A_{ij}}{\Delta} K h_j{\partial W
\over \partial x_j}\right)= \rho \Phi_{T0}
\label{eWeR}.\ee 
Noting that ${\cal A}=0,$ the unforced form of (\ref{eWeR})
gives  rotationally governed inertial modes in the core (see Papaloizou
and Pringle 1981, SPA). The local dispersion relation is (\ref{dispr})
with ${\cal A}=0.$ In order to solve (\ref{eWeR}) boundary conditions for
$W$ are needed. In the low frequency limit we may use the fact that 
the stratification in the radiative zone leads to $\xi_r$ being small
compared to the other components of the displacement. This is a familiar
property of oscillation modes such as $g$ modes in radiative layers.
This suggests the boundary condition $\xi_r=0$ at the convective core boundary.
Using (\ref{xii}), this can be expressed as a condition on $W.$

\noindent For real $\sigma$   finding $W$ in the convective core is problematic
because of the dense spectrum of the inertial modes (Greenspan 1968,
Papaloizou and Pringle 1981, SPA). However, if $\sigma$ is complex as
we shall assume, singular responses are avoided except possibly when
there are near resonances with global modes. A similar situation is expected
if viscosity is introduced. Solving (\ref{eWeR}) subject to $\xi_r=0$
on the convective core boundary is equivalent to solving an elliptic
boundary value problem and the solution will provide $W$ on the core
boundary which may be used as a boundary condition there for the problem of
finding $W$ in the radiative zone.

\noindent The elliptic boundary value problem described above is not simply
soluble in general, even in the low frequency limit. However, an exception
occurs when the core is assumed to have constant density and pressure,
an assumption which is only justifiable when the core is small.
Then a solution can be found in the form of a finite polynomial.
To find such solutions, it is best to write equation (\ref{eWeR}) 
in cylindrical polar coordinates $(\varpi,\varphi,z)$ in the form
\eq {\rho\sigma^2\over \Gamma
P}W +{1\over
(1-x^2)}\left[\frac{1}{\varpi}\frac{\partial}{\partial \varpi}
\left(\varpi\frac{\partial W}{\partial \varpi}\right)-{m^2 W\over
\varpi^2}\right] +\frac{\partial^2 W}{\partial z^2} ={3f\rho^2\sigma^2\over
\Gamma P^{(1+1/\Gamma)}}\varpi^2
\label{homc},\ee 
where
$$x=\frac{2\Omega_s}{\sigma}.$$
In order to proceed we take $P$ and $\rho$ to be constant and
equal to their values on the convective core boundary.
In the low frequency limit the first term on the left hand  side of
(\ref{homc}) may be neglected. The problem is then equivalent to
calculating the response of an homogeneous incompressible sphere
(see Greenspan, 1968). In this case the solution that has zero normal
displacement at the core boundary can be written as a polynomial in  $\varpi$
and $z$ for $m=2$ in the form
\eq W= a_c\varpi^4 +b_c\varpi^2+c_cz^2\varpi^2, \label{homW}\ee
where
$$a_c={3f\rho^2\sigma^2\over 2\Gamma P^{(1+1/\Gamma)}}
{(1-x^2)(2+x)\over (14-x^2+7x)},$$
$$b_c=-{3f\rho^2 r_c^2\sigma^2\over 2\Gamma P^{(1+1/\Gamma)}}
{(1+x)(4-x^2)\over (14-x^2+7x)},\ \ \ \ {\rm and}$$
$$c_c={3f\rho^2\sigma^2\over 2\Gamma P^{(1+1/\Gamma)}}
{(1+x)(2-x)\over (14-x^2+7x)},$$
where $r_c$ is the  radius of the convective core. We remark that the
response
is resonant when $14-x^2+7x=0.$ There are thus only two real resonant forcing
frequencies corresponding to the roots of the quadratic given by
$x=(7 \pm \sqrt{(105)})/2,$ or $\omega_f/\Omega_s = (\pm
\sqrt{(105)}- 7)/28.$ Although the spectrum is dense, the nature
of the forcing potential allows only two potential resonances to be excited
in this case. It would seem reasonable to assume that other modes
can only be weakly excited in more general cases. In fact the above solution for a uniform core may be
regarded as the first member of a sequence of higher order 
polynomial approximations to solutions in more general cases (see Papaloizou
and Pringle, 1981). Accordingly we shall use
(\ref{homW}) to determine $W$ on the core surface in the analysis presented
below.

\section {Asymptotic solution in the radiative zone}
The governing equation (\ref{eWeb}) developed for small $\sigma^2 /{\cal A} $
can be solved by utilising the fact that the operator on the left hand
side is separable (see SPA). We may write
\eq W=\sum^{\infty}_{n=0}w_n(r)X_n(\mu) \label{deco}.\ee
Here the $X_n$ are a set of orthogonal functions of $\mu$ over $(-1,1)$
so that the expansion coefficients are given by
 $$w_n =\frac{\int_{-1}^1 W X_nd\mu}{\int_{-1}^1 X^2_nd\mu}.$$
The $X_n$  satisfy  a second order ordinary differential equation defined
through the first order pair (see SPA)
\eq -\frac{\partial {\cal D}_n}{\partial \mu}
+\frac{mx \mu   {\cal D}_n}{(1-\mu^2)}-\frac{m^2
X_n}{(1-\mu^2)}=-\lambda_n X_n,\ee 
\eq  {\cal D}_n\left(1- x^2 \mu^2 \right)=
-(1-\mu^2)\frac{\partial X_n}{\partial \mu}
-mx\mu  X_n.\ee
The eigenvalues $\lambda_n$ are determined so as
to make $X_n$ regular at $\mu =\pm 1.$
Note that they are functions of the parameter $x$ and thus
the forcing frequency and
the stellar rotation rate. Note too that for potentials of the type we
consider here which are even functions of $\mu,$ we may restrict ourselves
to $\lambda_n$ such that $X_n, {\cal D}_n$ are even and odd functions
of $\mu$ respectively.
 
\noindent Using (\ref{deco}) we obtain
\eq O_{\perp}(W)=-\sum^{\infty}_{n=0}\lambda_n w_n(r)X_n(\mu) .\ee
Using the expansion given by equation (\ref{deco})
 and the orthogonality of the
$X_n,$ equation (\ref{eWeb}) gives

\eq w_n\left( \lambda_n -{\sigma^2\rho r^2\over\Gamma P}\right)
 +{\sigma^2
\over KP^{1/\Gamma} } \frac{\partial}{\partial r}\left(\frac{r^2 K
P^{1/\Gamma}}{{\cal A}} 
 {\partial w_n\over \partial r}\right)
 =-{ \sigma^2\rho r^2 S_n\over\Gamma P^{1 +1/\Gamma} }
\label{eWed},\ee
where $$S_n =\frac{\int_{-1}^1 SX_nd\mu}{\int_{-1}^1 X^2_nd\mu}.$$
The problem is thus reduced to solving a set of second order
ordinary differential equations for $w_n(r).$ This is possible because of the
fact that the normal mode problem becomes separable in the traditional
approximation (see Chapman and Lindzen, 1970 for a discusion in the context
of atmospheric tides and also SPA).
This approximation neglects the $\theta$ component of the stellar angular
velocity (Unno et al 1989), and it is expected to become valid in the stratified
radiative layers in the limit of low ferquencies when the radial displacement
becomes small compared to the angular displacements. 

\subsection{Reduction of S}
 From direct calculation using equation
(\ref{eWe}) we obtain for large ${\cal A}$
 \eq S={\Gamma P^{1-1/\Gamma}\over r^2}{\partial \over \partial r} \left(
{ r^2 P^{1/\Gamma}\Phi_{T0}\over g}\right) -{x^2\Gamma P\over gr}{\partial \over
\partial \mu} \left({\mu(1-\mu^2)\Phi_{T0}\over (1-x^2\mu^2)}\right)
+{mx \Gamma P\Phi_{T0}\over gr(1-x^2\mu^2)} +\rho\Phi_{T0},\ee
and after performing an integration by parts
\eq S_n \int_{-1}^1 X^2_nd\mu =\int_{-1}^1 SX_nd\mu =
f\rho r^2\left({\cal I}{\Gamma P^{1-1/\Gamma}\over\rho r^4}{d\over dr}
\left( { r^4 P^{1/\Gamma}\over g}\right)
-{x^2\Gamma P\over \rho g r}{\cal J}
+{\cal I}\right),\label{src}\ee
with $${\cal I}=\int^1_{-1}X_n P^2_2 (\mu)d\mu,$$ and
$${\cal J}=\int^1_{-1}(\mu{\cal D}_n - mX_n/x) P^2_2 (\mu)d\mu.$$

\subsection{Behaviour of the eigenvalues $\lambda_n(x)$}
The calculation of the tidal
torque has thus been reduced to solving the sets of second order
differential equations (\ref{eWed}). The boundary conditions are that the
solution should, if possible, correspond to a predominantly outgoing wave at
large radii, and $w_n$ should be matched at the core boundary to the
solution obtained above that applies to the convective core. 

\noindent Clearly the form of the solutions of the determining
equation (\ref{eWed}) depends on the eigenvalues $\lambda_n.$
>From general
scaling arguments one expects $w_n = O(\sigma^2/\lambda_n)$ which vanishes as
$\sigma$ vanishes. When $\lambda_n \gg 0,$ the unforced solutions for $w_n$ are 
short wavelength $g$ mode like waves. There is special interest in
the smallest positive $\lambda_n$ because these lead to solutions
with the longest radial wavelengths which will give the strongest
responses to global forcing of the type considered here.

\noindent For a non-rotating star $( x=0)$  the eigenvalues $\lambda_n =l(l+1),$
with $l\equiv n = 2,4,6...$ and the corresponding eigenfunctions $X_n=
P^2_l(\mu).$ However, the behaviour of $\lambda_n$ is very different in the
inertial regime for which $-1<x^{-1}<1.$ We plot the smallest
positive $\lambda_n$ we obtained for a range  of values of $x$ in the inertial
regime in figure 1. It will be seen that there is a significant difference
between positive and negative $x.$ For $ x>0,$ corresponding to prograde
relative rotation of the companion, $\lambda_n$ is similar to the non-rotating
case. However, for $x<0,$ corresponding to retrograde relative rotation,
$\lambda_n$ becomes large for $x^{-1}<-1/6.$ These eigenvalues correspond to
disturbances confined cosely to the equator. For $-1/6<x^{-1} <0,$ another small
$\lambda_n$ exists corresponding to a toroidal mode resonance.
These eigenvalues are important because they lead to solutions
with strong global responses to the forcing tide.

\subsection{Toroidal mode resonances}
Strict toroidal mode resonance can be defined to occur  when $x$ is such
that an eigenvalue $\lambda_n =0,$ (see SPA). Then we have
$x=-(l(l+1))/m,$ with $l\equiv n \ge |m|$ being an odd integer to provide
solutions with the required symmetry type. These resonances
corespond to the $r$ mode frequencies (Papaloizou and Pringle, 1978)
$$\sigma ={-2m\Omega_s\over l(l+1)}.$$
The corresponding eigenfunctions  are ${\cal D}_n= P^{|m|}_l(\mu),$ and  
$X_n =m^{-2}(-l(l+1) \mu P^{|m|}_l(\mu)-(1-\mu^2)dP^{|m|}_l(\mu)/d\mu),$ where
$P^{|m|}_l(\mu)$ denotes the standard Legendre function. The lowest order
resonance has $l=3$ so that $x^{-1} =-1/6.$

\noindent We see that, for  positive values of $m$ , the toroidal
mode resonances occur for negative forcing frequencies, corresponding to
retrograde orbital rotation relative to the star. In this case the  star
rotates faster than the orbit so that the action of the tides is to cause
the star to spin down. It is clear from the above discussion that the
existence of toroidal $r$ mode resonances leads to a qualitatively different
tidal response for some negative forcing frequencies which is in general
much larger than for corresponding positive forcing frequencies. The
tidal evolution timescale will accordingly be shorter in that case.

\section{WKB solution}
It is possible to solve equation (\ref{eWed}) by a Green's
function method as in Papaloizou and Savonije (1985). In general
extended regions outside the convective core are found to contribute
to the excitation of an outgoing wave. However, for
simplicity we here look at the solution of equation (\ref{eWed}) in the low
frequency limit where a WKB approximation should be adequate. In this case the
largest lenghthscale associated with the solution for $w_n$ occurs near the
convective core boundary (Zahn, 1977) where, for $\lambda_n > 0,$ an
outgoing wave is excited. Only this region matters for wave excitation in
the limit of low forcing frequencies but if this limit is to be of practical
use, the convective core should not be too small. 

\noindent To construct the solution, we write $z =r-r_c,$ with $r_c$
being the radius of the convective core boundary. Further we assume a local
first order Taylor expansion such that ${\cal A} =
({\cal A}')_c z,$ and evaluate all other quantities in  (\ref{eWed}) at the core
boundary. We then obtain
\eq w_n\left( \lambda_n -\left({\sigma^2\rho r^2\over\Gamma P}\right)_c\right)
 +\left({\sigma^2r^2
\over {\cal A}' }\right)_c \frac{\partial}{\partial z}\left({1\over z}
 {\partial w_n\over \partial z}\right)
 =-\left({ \sigma^2\rho r^2 S_n\over\Gamma P^{1 +1/\Gamma} }\right)_c
\label{eWef}.\ee
Here the subscript $c$ denotes evaluation at the core boundary. From 
now on we shall take this as read for all state variables and the subscript  $c$
will be dropped.

\noindent It is straightforward to express the solution of
(\ref{eWef}) that coresponds to only outward going waves in terms of Hankel
functions, $H^{(1)}_{\nu},$  as follows. If $Q=z^{-1}dw_n/dz,$ then the solution
for $Q$ is given by \eq Q= {3^{5/6}\Gamma(4/3)\exp(-i\pi/6)\over
2a^{1/6}}C_0z^{1/2}H^{(1)}_{1/3} \left({2\over 3} a^{1/2} z^{3/2}\right),\ee
where \eq a={{\cal A}'\over \sigma^2 r^2}\left(\lambda_n-{\sigma^2\rho
r^2\over \Gamma P}\right), \label{nn}\ee
\eq C_0=-a w_n(0)-{S_n\rho{\cal A}'\over\Gamma P^{1+1/\Gamma}}, \label{fg}\ee
with $w_n(0)$ denoting the prescribed value of $w_n$ on the core boundary.

\noindent Here it is assumed that the real part of the forcing frequency is
positive. Solutions appropriate to negative forcing frequencies can be found
by setting $m\equiv 2\rightarrow -m,$  below while retaining the forcing frequency
as positive.

\noindent The $w_n(0)$ are given by
$$w_n(0) =\frac{\int_{-1}^1 W X_nd\mu}{\int_{-1}^1 X^2_nd\mu},$$
with $W$ being evaluated on the core boundary. Using  the solution given by
(\ref{homW}), we obtain
$$w_n(0) =\frac{\int_{-1}^1(a_
cr^4(1-\mu^2)^2
+b_cr^2(1-\mu^2)+c_cr^4\mu^2(1-\mu^2)) X_nd\mu}{\int_{-1}^1 X^2_nd\mu},$$ 
\subsection{Angular momentum flux and tidal torque}
The asymptotic outgoing wave solutions outlined above are associated
with a conserved angular momentum flow or wave
action ( Goldreich an Nicholson, 1989). Assuming the outgoing waves are
ultimately absorbed near the surface, this angular momentum is deposited there
and correspondingly removed from the orbit. But note that this exchange can be
of negative sign for retrograde forcing. Thus in this picture, the tidal torque
acts through the production and absorbtion of angular momentum carrying waves.

\noindent The radial component of the wave angular momentum flux appropriate for
responses to the complex forcing potential is given by
$$F = {m\over 2} {\cal IM}\left(P' \xi_r^{\ast}\right),$$
 where ${\cal IM}$ denotes the imaginary part
(see Ryu and Goodman, 1992, Lin et al, 1993).
Using 
$$\xi_r=-{K\over {\cal A}'}QX_n,\ {\rm and} \  P' =w_n X_ nP^{1/\Gamma}$$
in the asymptotic limit, we find the asymptotic form ( for large $z$ )
of the angular momentum flux associated with each $\lambda_n$
\eq F= {3^{8/3}\left(\Gamma(4/3)\right)^2Km P^{1/\Gamma}\sigma^2
r^2|C_0|^2 X_n^2 \over 8\pi
a^{1/3}({\cal A}')^2 \left( \lambda_n -\left({\sigma^2\rho r^2\over\Gamma
P}\right)\right)}\label{ang}.\ee
The total rate of change of angular momentum of the star is found
by integrating the radial flux over a spherical surface at the convective
core boundary. It is thus
given by
$${\dot J}= 2\pi r^2\int^{1}_{-1}Fd\mu.$$

\noindent Using this, while neglecting
$(\sigma^2\rho r^2)/(\Gamma P)$ in comparison to 
$\lambda_n,$  we find after some algebraic manipulation that for a particular
$\lambda_n,$ 
\eq {\dot J}= \rho r^5
\epsilon_T \Omega_c^2 {3^{8/3}\left(\Gamma(4/3)\right)^2\Theta^2|\Xi|^2\over
2^{7/3} \lambda^{4/3}_n I_n}\left({|\omega_f|\over \Omega_c}\right)^{8/3}
\left({\Omega_c^2\rho r^2\over \Gamma P}\right)^{2} \left({\Omega_c^2\over
r{\cal A}'}\right)^{1/3},\label{crp}\ee
Here $$\epsilon_T ={M^2_pR_s^6\over M_s^2 D^6},$$ and $\Theta$ is defined
through $$S_n = -{f\rho r^2\over I_n}\left[{\cal I}\left({\Gamma P\over \rho g
r} \left({4\pi\rho r^3\over M}-6\right)\right)+
{\cal J}{x^2\Gamma P\over \rho g r}\right]\equiv {f\rho r^2\over
I_n} \Theta,$$ with  $M$ being the mass interior to radius $r$ and
$$I_n =\int^{1}_{-1}X_n^2 d\mu.$$
The quantity $\Xi$ is defined by
$$\Xi=1+{w_n(0)\lambda_n\Gamma P^{1+1\Gamma}I_n\over
f\rho^2r^4\sigma^2\Theta},$$ and $\Omega_c^2=GM_sR_s^{-3}.$

\noindent Here we have set $m=2,$ corresponding to prograde forcing frequencies.
In this case ${\dot J} > 0,$ corresponding to stellar spin up.
For negative forcing frequencies the same expression may be used but
the sign of ${\dot J}$ is reversed. In general we should sum over $\lambda_n.$
However, for  simplicity we shall restrict ourselves to the values displayed
in figure 1 for the numerical work described below. 

\noindent In order to compare with the work of
SP, we introduce $t_0$ defined through
$$ t_0= {kM_sR_s^2 \epsilon_T |\omega_f|\over {\dot J}},$$
with $kM_sR_s^2$ being the stellar moment of inertia.
Thus
\eq t_0 ={kM_sR_s^2  \over \Omega_c\rho r^5}
{ 2^{7/3}\lambda^{4/3}_n I_n\over3^{8/3}
\left(\Gamma(4/3)\right)^2\Theta^2|\Xi|^2}
\left({|\omega_f|\over \Omega_c}\right)^{-5/3}
\left({\Omega_c^2\rho r^2\over \Gamma P}\right)^{-2}\left({\Omega_c^2\over
r{\cal A}'}\right)^{-1/3} .\label{to}\ee

\noindent We remark that for a particular stellar model $\Theta^2|\Xi|^2=F(x)$
depends on the forcing and 
stellar rotation frequencies through the quantity
$x$ only. Thus we have the scaling law that
\eq t_0 \propto \left(|\omega_f|\right)^{-5/3}/F(x),\label{sca}\ee
which, as long as our simple prescription for including damping at inertial
core mode resonances is used ( see below ),  enables results to be scaled to
different forcing frequencies for the same value of $x.$ 

\section{Numerical
evaluation} 
We have evaluated $t_0$ using (\ref{to}) for the 20
$M_{\odot}$ stellar model considered in SP. We display the result for
$\log_{10}t_0$ plotted against $\omega_f/\Omega_s$ for
$\Omega_s =0.1 \Omega_c$   in the inertial regime in figure 2.
But note that these results may be scaled to other values of $\Omega_s$
using the scaling relation (\ref{sca}).

\noindent Although we do not reproduce all the details found in SP,
the most important features where they can be compared, including the form of the difference
between prograde and retrograde forcing, and agreement with the
 values of $t_0,$  to  order of magnitude,
 are found. This type of  qualitative agreement is reasonable in view of the large
amount of variation in $t_0$ and 
the fact that the asymptotic analysis is carried out in the limit of 
both low forcing and rotation  frequencies.
This results in  extreme sensitivity to the location of the convective core boundary.
 Also  simplifying approximations  were made to find the response of  the convective core. 
Note too that the assumption of an
outward going wave condition in the low frequency limit has prevented the
appearance of the oscillatory behaviour found in SP.  It is  hoped that a  future extension
of the analysis will enable relaxation of these approximations. 

  We  do find a strong
toroidal mode resonance at the expected location $\omega_f = - \Omega_s/6.$ 
This occurs because of the long wavelength response near this location
resulting from the small $\lambda_n.$ We remark that the assumption
of an outgoing wave condition will break down in the centre of this resonance
where the response is of very long wavelength. Also we find rather small values
of $t_0$ near $x^{-1}=0$ ( note the resolution here was $0.01$)
 in a domain that could not be considered in SP. The origin of this behaviour can be traced 
to the forcing term $\propto x^2$ in (\ref{src}).  

\noindent As  in SP, we obtain
larger $t_0$ in general for  $\omega_f <  - \Omega_s/6$ because of the larger
$\lambda_n.$ Note too that we have two core inertial mode resonance
features at $\omega_f \sim - 0.6 \Omega_s$ and $\omega_f \sim  0.12 \Omega_s.$
Similar features were seen in SP. To obtain features of similar scale, we
incorporated damping in the core by adding a negative imaginary part to 
$\omega_f $ of magnitude $0.06|Re(\omega_f)|$ when calculating $w_n(0).$ In reality the damping is
likely to be  due to $g$ mode losses, that would appear in an improved higher order
theory, as well as applied viscosity. With our
simple damping prescription, the details of these particular features are not
too well reproduced but this is not too surprising in view of the simplifying
assumption of an homogeneous core.

\section{Conclusions}
We have developed a generalization of the asymptotic treatment
of the tidal torques experienced by a massive star as a result of a companion
in circular orbit considered by Zahn (1977) and Papaloizou and Savonije (1985)
to the case of a rotating star when the forcing frequency is less than the
rotation frequency in magnitude. The results confirm the presence of
a strong toroidal mode resonance feature for retrograde forcing and also,
albeit with a very simplified model of the convective core,
the presence of some core inertial mode features found in SP.
\newpage

\begin{figure}
\mbox{\epsfxsize=8.0 cm
     \epsfysize=10.0 cm
      \rotate[r]{\epsffile{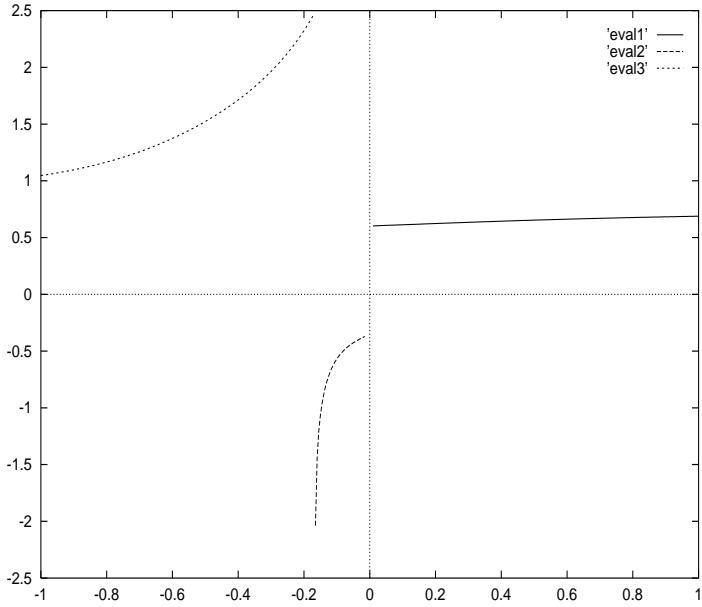}}}
\caption[] {The form of $\log_{10}\lambda_n$ as a function of $x^{-1}$
used in the evaluation of $t_0.$ These $\lambda_n$ represent the lowest
positive values outside the interval (-1/15,0)}
\label{fig1}  
\end{figure}

\begin{figure}
\mbox{\epsfxsize=8.0 cm
     \epsfysize=10.0 cm
      \rotate[r]{\epsffile{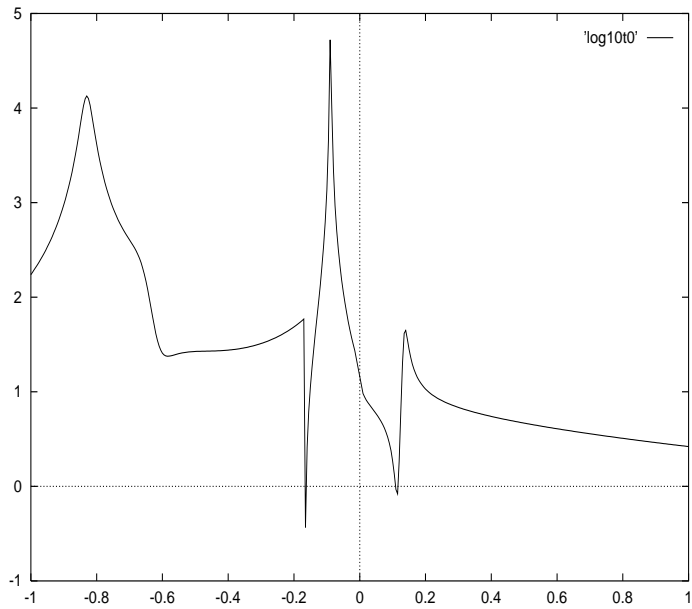}}}
\caption[] {A plot of $\log_{10}t_0,$ $t_0$  in yr.,  as a function of $x^{-1}$
in the inertial regime for $\Omega_s = 0.1\Omega_c.$}
\label{fig2}  
\end{figure}

\end{document}